\documentclass{emulateapj}
\usepackage{amsmath} 
\usepackage{graphicx}
\usepackage{natbib}
\usepackage{bm}
\usepackage[caption=false]{subfig}
\usepackage{tabularx}
\usepackage{booktabs}
\usepackage{longtable}

\newcommand{\q}{\mathbf{q}}
\newcommand{\x}{\mathbf{x}}

\newcommand{\s}{\mathbf{s}}
\newcommand{\nablab}{\mathbf{\nabla}}

\begin{document}

\title{Nonlinear Behavior of Baryon Acoustic Oscillations in Redshift Space from the Zel'dovich Approximation}
\author{Nuala McCullagh and Alexander S. Szalay}
\affil{The Johns Hopkins University \\ Baltimore, MD 21218}

\begin{abstract}
Baryon acoustic oscillations (BAO) are a powerful probe of the expansion history of the universe, which can tell us about the nature of dark energy. In order to accurately characterize the dark energy equation of state using BAO, we must understand the effects of both nonlinearities and redshift space distortions on the location and shape of the acoustic peak. In a previous paper we introduced a novel approach to 2nd order perturbation theory in configuration space using the Zel'dovich approximation, and presented a simple result for the first nonlinear term of the correlation function. In this paper, we extend this approach to redshift space. We show how perform the computation, and present the analytic result for the first nonlinear term in the correlation function. Finally, we validate our result through comparison to numerical simulations.
\end{abstract}


\section{Introduction}
\label{sec:intro}

In the early universe, when photons were tightly coupled to baryons, the gravitational collapse of baryonic density perturbations was opposed by photon pressure, giving rise to acoustic waves in the baryon-photon plasma \citep{peeblesyu1970,silk1968, sz1970, bondefstathiou1984}. The signature of this effect is known as Baryon Acoustic Oscillations (BAO), and is observable in both the temperature power spectrum of the cosmic microwave background (CMB) \citep{debernardis2000,hanany2000}, and the statistics of the matter and galaxy distributions. The BAO feature was first detected as a peak in the correlation function of the Luminous Red Galaxies (LRGs) in the Sloan Digital Sky Survey (SDSS) by \citet{eisenstein2005}, and was subsequently measured in several other galaxy surveys \citep{cole2005, padmanabhan2007, percival2007, percival2010, blakeBAO2011}. 

The BAO signal is a useful cosmological probe because even at low redshift, the scale of interest is still largely in the linear regime. The location of the peak thus acts as a ``standard ruler'' with which the expansion history of the universe can be measured as a function of redshift. This allows us to constrain cosmological parameters, including the equation of state of dark energy \citep{seoeisenstein2003}.

In order to accurately constrain the properties of dark energy using BAO, it is important to model the systematic uncertainty of the standard ruler. Linear theory, which predicts that the amplitude of the peak will scale as the square of the linear growth function, $D(t)$, does not take into account the nonlinear growth of structure, which causes the position and strength of the acoustic peak to change over time.  Also, because we are forced to use redshifts to determine distances along the line of sight, we observe an anisotropic galaxy distribution instead of the expected isotropic distribution. The linear order effects of redshift-space distortion is well understood \citep{kaiser1987, tian2011, blake2011, seo2012}. Understanding how these distortions affect higher order terms will be increasingly important as observations become more precise.

A great deal of work has been done to characterize the nonlinearities using cosmological perturbation theory \citep{jainbertschinger, bouchet1995,bernardeau2002,croccescoccimarro2006, croccescoccimarro2008, matsubaraRPT2008}, and to extend perturbation theory into redshift space \citep{matsubaraPTRSD2008, taruya2010, valageas2011, carlson2013}. Much of this previous work has been done almost exclusively in Fourier space, focusing on the power spectrum. However, configuration space has many advantages over Fourier space when analyzing the BAO signal. Observationally, it is more natural to work in configuration space, as the BAO signal is a localized peak in the correlation function. Also, in Fourier space, the higher order terms from perturbation theory involve convolutions over complicated kernels. In configuration space, these terms may involve only sums of products of linear correlation functions. Finally, because redshift space is a simple transformation from configuration space, the full redshift-space correlation function can be straight-forwardly computed to any order.

There has been renewed interest recently in using the Zel'dovich approximation to model the nonlinear evolution of dark matter \citep{mccullagh2012, carlson2013, white2014, tassev2014a, tassev2014b}. On BAO scales, the Zel'dovich approximation may provide higher accuracy with relatively less computational difficulty than the Fourier-space approach of standard perturbation theory \citep{white2014, tassev2014a}. We have previously shown that one can calculate the real-space correlation function to second order from the Zel'dovich approximation, and that the resulting expression agrees with the previously calculated nonlinear power spectrum \citep{mccullagh2012, grinsteinwise1987}. In this paper, we extend our configuration-space approach to perturbation theory in order to calculate the nonlinear correlation function in redshift space from the Zel'dovich approximation. In Section~\ref{sec:first}, we derive the nonlinear term in the redshift-space correlation function analytically starting from the Zel'dovich approximation and using the distant observer approximation to relate real to redshift space. We show in Section~\ref{sec:discussion} that the term is in agreement with numerical simulations of the Zel'dovich approximation. We conclude in Section~\ref{sec:conclusion}.

\section{Theory}
 \label{sec:first}
 
In the Zel'dovich approximation, particles' initial Lagrangian coordinate, $\q$, and their co-moving Eulerian coordinate, $\x$, are related through a displacement field $\bm{\psi}$. The displacement field is the gradient of the linear potential, $\phi(\q)$, times the growth function $D(t)$ \citep{zeldovich1970, shandarin1989}.  The the linear potential is related to the linear overdensity field through the Poisson equation:
\begin{align}
\x(\q, t) &= \q+\bm{\psi}(\q,t)\ ,\label{eq:Zel'dovich}\\
\nablab_q\cdot\bm{\psi}(\q,t)&= -D(t)\nabla_q^2\phi(\q)=-\delta_{L}(\q,t)\ .\notag
\end{align}

To a distant observer, the redshift coordinate $\s$ is related to the Eulerian coordinate $\x$ through the particle's peculiar velocity along the line of sight:
\begin{align}
\s(\x)&=\x+f(\bm{\psi}(\q,t)\cdot \hat n)\hat n\ ,\label{eq:redshiftspace}
\end{align}
where $\hat n$ is the line-of-sight direction, and $f$ is the logarithmic derivative of the growth function, $D$ (see \citet{hamilton1998} for a review). Combining Equations (\ref{eq:Zel'dovich}) and (\ref{eq:redshiftspace}), we write the transformation from Lagrangian to redshift coordinates in the Zel'dovich approximation in a similar way to Equation (\ref{eq:Zel'dovich}), but with an anisotropic displacement field:
\begin{align}
\s(\q)&=\q+\bm{\Psi}(\q,t)\label{eq:l2rs}\\
\mathbf{\Psi}(\q,t)&\equiv-D(t)\left( \nablab_q \phi(\q)+f(\nablab_q\phi(\q)\cdot\hat n) \hat n\right)\notag
\end{align}
 
We now follow the same approach outlined in \citet{mccullagh2012} to calculate the nonlinear overdensity, but using the anisotropic displacement field. Mass conservation allows us to relate the initial Gaussian overdensity to the final redshift-space overdensity through the (anisotropic) Jacobian of the transformation in Equation (\ref{eq:l2rs}): 
 \begin{align}
\frac{\rho(\s, t)}{\bar \rho}&=\left | \frac{\partial s_i}{\partial q_j} \right |^{-1}=\frac{1}{J(\q, t)} =  1 + \delta_q(\q(\s))\ ,
\label{eq:euleriandensity}
\end{align}
 where $\s$ and $\q$ are related by Equation (\ref{eq:l2rs}). Here, $\delta_q$ (without subscript $L$) is the {\em weakly nonlinear} over-density in redshift space, expressed through the Lagrangian coordinate $\q$. 
 
 As discussed in the previous paper, Equation (\ref{eq:euleriandensity}) is only strictly valid before shell-crossing, where the mapping is one-to-one. We make the assumption that on the scales we are interested in, the effects of shell-crossing are negligible.
 
The Jacobian of the transformation from $\q$ to $\s$ can be computed using the relation in Equation \ref{eq:l2rs}. It can be expressed in terms of $f$ and elements of the symmetric deformation tensor $d_{ij}(\q)$, which is the second derivative of the isotropic gravitational potential \citep{zeldovich1970}:
 \begin{align}
d_{ij}(\q)&=\frac{\partial^2 \phi(\q)}{\partial q_i \partial q_j}\ ,
\end{align}
\begin{align}
J(\q, t)&=1-D (I_1(\q)+fd_{nn}(\q))+D^2 \big( (1+f)I_2(\q)\notag\\
&\qquad-f M_{nn}(\q)\big)-D^3 (1+f)I_3(\q)\ ,
\end{align}
where $d_{nn}$ is the line-of-sight component of $d_{ij}$, $M_{nn}$ is the corresponding minor of the matrix, and $I_1$, $I_2$, and $I_3$ are invariants of the matrix, which can be written in terms of its eigenvalues $\lambda_1$, $\lambda_2$, and $\lambda_3$:
\begin{align}
I_1(\q)&=\lambda_1+\lambda_2+\lambda_3\ ,\notag\\
I_2(\q)&=\lambda_1\lambda_2+\lambda_1\lambda_3+\lambda_2\lambda_3\ ,\\
I_3(\q)&=\lambda_1\lambda_2\lambda_3\ .\notag
\end{align}

It is necessary to express the overdensity in terms of the redshift coordinate, rather than the Lagrangian coordinate, because we want the expression for the correlation function in terms of the observed redshift-space separation, $\s_1-\s_2$. To do this, we first invert Equation (\ref{eq:l2rs}):
 \begin{align}
 \q(\s)&=\s-\mathbf{\Psi}(\q(\s),t)\ ,
 \end{align}
and use Taylor expansions of relevant quantities about the point $\q=\s$. First, we express $\mathbf{\Psi}$ as a function of redshift coordinate:
\begin{align}
\Psi_{i}(\s(\q))&=\left(\Psi_{i}(\q)-\sum_j \frac{\partial \Psi_{i}(\q,t)}{\partial q_j}\Psi_{j}(\q,t)+... \right)\Bigg|_{\q=\s}\ ,
\end{align}
where $\Psi_i$ indicates the $i$-th component of $\mathbf{\Psi}$. Because $\mathbf{\Psi}$ is proportional to $D$, this expansion is a power series in $D$. 

Next, we want the final redshift-space overdensity in terms of $\s$, which we will call $\delta_s(\s)$. This function is equivalent to the overdensity $\delta_q$ found in Equation \ref{eq:euleriandensity}, evaluated at the corresponding Lagrangian coordinate, $\q(\s)$:
 \begin{align}
 \delta_s(\s,t)&\equiv\delta_q(\q(\s),t)\label{eq:densityqrs}
 \end{align}
 In order to find the functional form of $\delta_s(\s)$, we rewrite the right-hand side of Equation \ref{eq:densityqrs} as:
 \begin{align}
 \delta_q(\q(\s),t)&=\delta_q(\s-\mathbf{\Psi}(\q(\s)))\label{eq:taylordensity}.
 \end{align}
A Taylor expansion of the right hand side of Equation (\ref{eq:taylordensity}) gives:
\begin{align}
\delta(\s,t)&=\bigg(\delta(\q,t)-\sum_i \frac{\partial \delta(\q)}{\partial q_i}\Psi_{i}(\q(\s)) \notag\\
&\qquad+ \frac{1}{2}\sum_{i,j}\frac{\partial \delta(\q)}{\partial q_i\partial q_j} \Psi_{i}(\q(\s))\Psi_{j}(\q(\s))+...\bigg)\bigg|_{\q=\s}
\end{align}

The full expression for the redshift-space over-density in terms of the Lagrangian over-density to third order in $D$ is:
\begin{align}
	\delta(\s,t)&=\bigg(\delta(\q,t) - \sum_i \frac{\partial\delta(\q,t)}{\partial q_i}\Psi_{i}(\q)\notag\\
	&\qquad+\sum_{i,j}\frac{\partial \delta(\q,t)}{\partial q_i}\frac{\partial\Psi_{i}(\q,t)}{\partial q_j}\Psi_{j}(\q,t)  \notag\\
	&\qquad+ \frac{1}{2} \sum_{i,j}\frac{\partial^2\delta(\q,t)}{\partial q_i \partial q_j}\Psi_{i}(\q)\Psi_{j}(\q)+... \bigg)\Bigg|_{\q=\s}\ .
	\label{eq:rsdensity}
\end{align}
In this expression, $\delta(\q, t)$ is the Lagrangian overdensity computed from the inverse Jacobian, as in Equation \ref{eq:euleriandensity}. This is a function of the invariants of the deformation tensor, which are related to the linear over-density $\delta_L$. The quantity $\Psi(\q, t)$ in Equation \ref{eq:rsdensity} is simply the gradient of the linear potential, which is related to the linear overdensity through Poisson's equation (Equation \ref{eq:Zel'dovich}). Thus we have an expression for the nonlinear Eulerian overdensity in redshift space as a function of the linear overdensity and its derivatives. This can now be used to compute the correlation function in redshift space.  

\begin{figure}
\begin{center}
\includegraphics[scale=0.45]{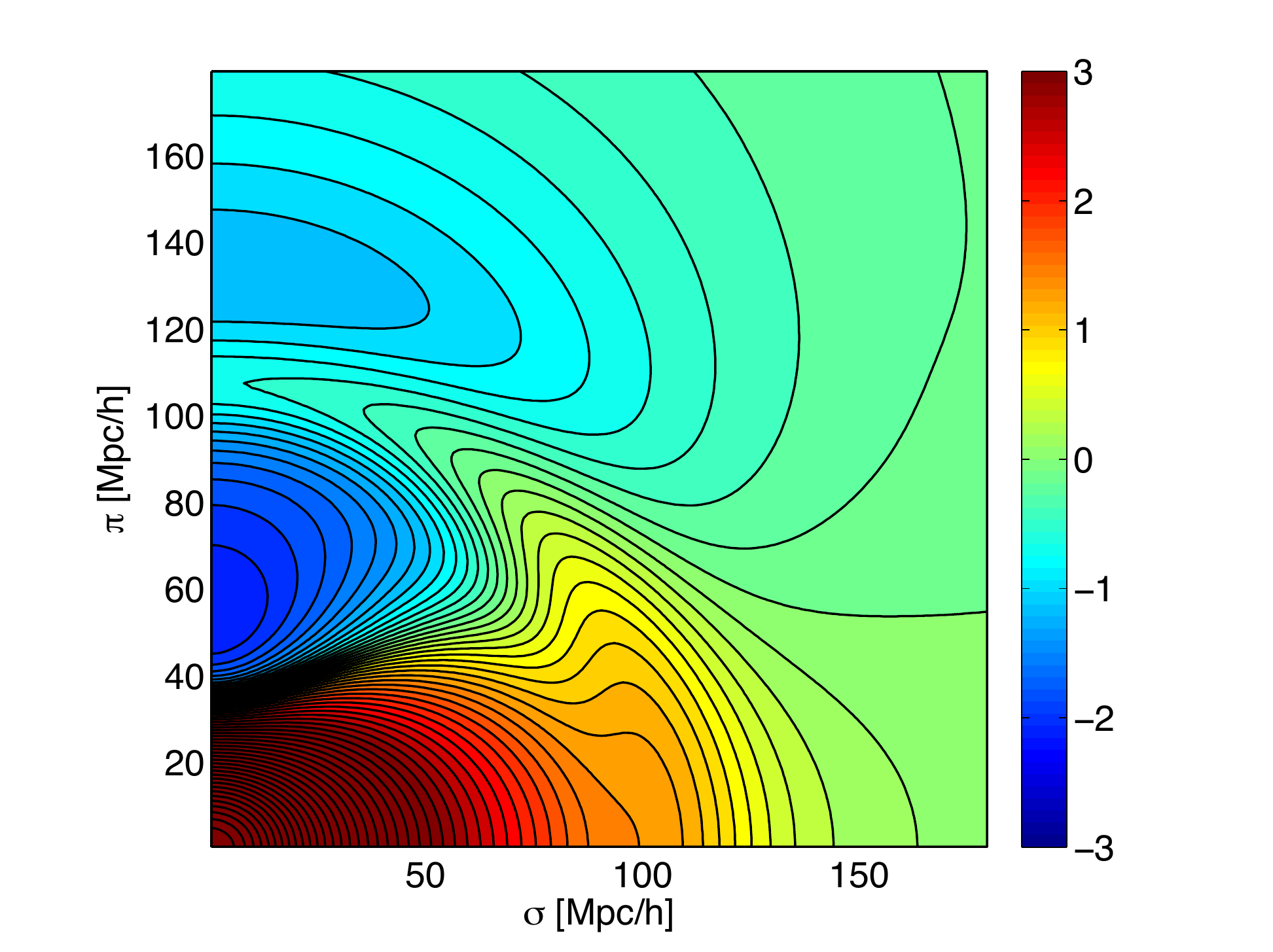}
\caption{The two-dimensional linear correlation function in redshift space. We plot $\sinh^{-1}(500\xi^{(1)}(\s))$, which is linear near zero and logarithmic for high values of $\xi$.}
\label{fig:cflin}
\end{center}
\end{figure}

In co-moving coordinates, the redshift-space correlation function is defined as:
\begin{align}
\xi(\s,t)&\equiv\langle \delta(\s',t)\delta(\s'+\s,t)\rangle\ .
\end{align}
Plugging in the density from Equation \ref{eq:rsdensity} and grouping terms by order in $D$, we arrive at a perturbative expression for the correlation function that is a function only of the linear density and its derivatives. This can be written as:
\begin{align}
\xi(\s,t)&=D^2\xi^{(1)}(\s,t)+D^4\xi^{(2)}(\s,t) \ ,
\end{align}
where the terms proportional to odd powers of $D$ vanish because $\delta_L$ is assumed to be a Gaussian random field. The $\xi^{(1)}$ term is simply the variance of linear density ($\delta^{(1)}$) at separation $\s$. $\xi^{(2)}$ contains two terms, one the expectation value between $\delta^{(2)}$ and itself, and the other the expectation value of $\delta^{(1)}$ and $\delta^{(3)}$ at separation $\s$.
 
As in real-space, the redshift-space correlation function can be written in terms of the linear correlation functions:
\begin{align}
\xi_n^m(s)&=\frac{1}{2\pi^2}\int_0^{\infty} P_L(k)j_n(ks)k^{m+2}dk \ .
\end{align}
These functions are 1-dimensional integrals of the linear power spectrum, and thus can be computed very quickly for each combination of $n$ and $m$.

The calculations of the expectation values in redshift space are very similar to those in real space, but now involve the anisotropic displacement vector. For example, the expectation value between the line-of-sight component of the displacement vector and its line-of-sight derivative is:
\begin{align}
\Big \langle \Psi_n(\s_1)& \frac{\partial \Psi_n(\s_2)}{\partial q_n} \Big \rangle=\notag\\
&(1+f)^2 \left(\frac{2}{5}\mathcal{P}_3(\mu)\xi_3^{-1}(s) - \frac{3}{5} \mathcal{P}_1(\mu)\xi_1^{-1}(s)\right)
\end{align}
In real space, the result would be the same, but without the factor of $(1+f)^2$. See \citet{mccullagh2012} for a more detailed description of how to calculate these expectation values using spherical harmonics. 

\begin{figure}
\begin{center}
\includegraphics[scale=0.45]{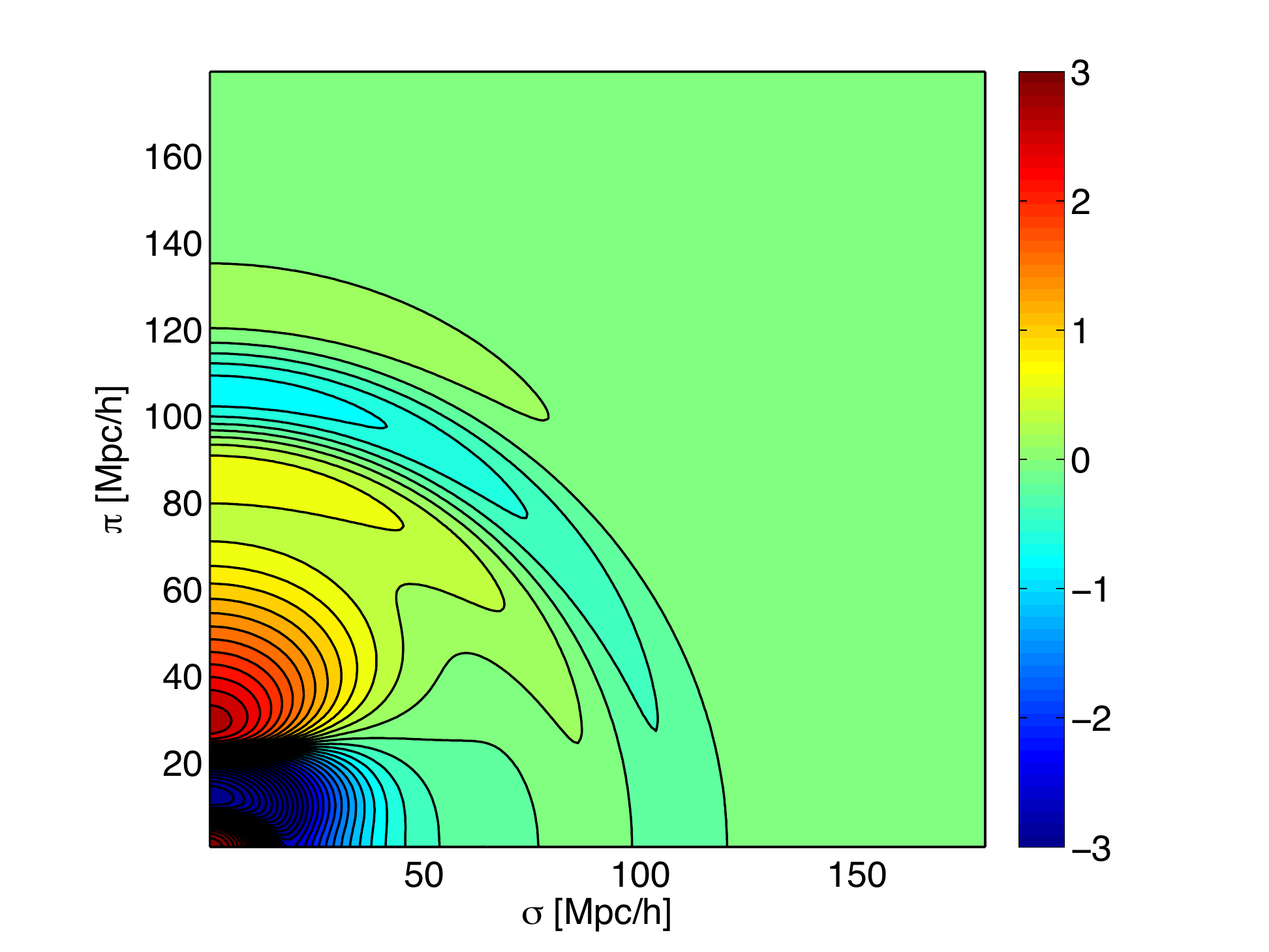}
\caption{The second order term of the Zel'dovich correlation function in redshift space. Again, we plot $\sinh^{-1}(500\xi^{(2)}(\s)$.}
\label{fig:cfnonlin}
\end{center}
\end{figure}

The linear redshift-space correlation function, $\xi^{(1)}(\s)$, can be computed in this way:
\begin{align}
\xi^{(1)}(s, \mu)&=\left (1+\frac{2 f}{3}+\frac{f^2}{5} \right)\xi_0^0(s)\notag \\
&-\left(\frac{4f}{3}+\frac{4f^2}{7}\right)\mathcal{P}_2(\mu)\xi_2^0(s)+\frac{8f^2}{35}\mathcal{P}_4(\mu)\xi_4^0(s)\label{eq:cflin}
\ ,
\end{align}
where $\mathcal{P}_{\ell}(\mu)$ is the Legendre polynomial of order $\ell$. This is the configuration-space equivalent of the linear Kaiser formula \citep{kaiser1986, hamilton1992, hamilton1996}. 

As in real space, the first nonlinear term contains expectation values of four linear quantities, which simplifies to products of the functions $\xi_n^m(s)$ due to Wick's theorem. In redshift space, the term has additional dependence on $f$ and $\mu$, as expected. We have condensed the terms for convenience, and give the full expression in the Appendix.
\begin{align}
\xi^{(2)}(s, \mu)&=\sigma_v^2\sum_{\ell=0}^{4}f^{\ell/2}B_\ell(f)\xi_{\ell}^2(r)\mathcal{P}_\ell(\mu) \notag \\
&+ \sum_{\ell=0}^8\sum_{n_1, n_2}^6 \sum_{m=0}^2 f^{\ell/2}A_{n_1, n_2}^{\ell, m}(f) \xi_{n_1}^m(s) \xi_{n_2}^{-m}(s) \mathcal{P}_{\ell}(\mu)\label{eq:xi2rs}
\end{align}
where
\begin{align}
\sigma_v^2\equiv \frac{1}{6\pi^2} \int_0^{\infty} P_L(k) dk
\end{align}

There are several interesting cases we can study using this expression. The azimuthally averaged correlation function, which uses a $\sin \theta \ d\theta$ weighting, picks out only the $\ell=0$ terms from the anisotropic correlation function. This is due to the orthogonality of the Legendre polynomials. The linear contribution to the azimuthally averaged correlation function is $\bar{\xi}^{(1)}(s)=(1+\frac{2}{3}f+\frac{1}{5}f^2)\xi_0^0(s)$. Using equation \ref{eq:xi2rs}, we can easily compute the nonlinear contribution to the averaged correlation:
\begin{align}
\bar{\xi}^{(2)}(s)&=\sigma_v^2 B_0\xi_{0}^2(r)+A_{00}^{00}\xi_0^0(s)^2+A_{22}^{00}\xi_2^0(s)^2 \notag \\
&+A_{44}^{00}\xi_4^0(s)^2+A_{11}^{01}\xi_1^{1}(s)\xi_1^{-1}(s)+A_{33}^{01}\xi_3^{1}(s)\xi_3^{-1}(s)\notag \\
&+A_{22}^{02}\xi_2^{2}(s)\xi_2^{-2}(s)
\end{align}

Similarly, the quadrupole term can be computed by choosing only the $\ell=2$ terms, and so on. The full expressions for the nonlinear monopole and quadrupole are given in the Appendix.

\section{Discussion}
\label{sec:discussion}

In our previous paper, in order to verify our real-space result for the nonlinear correlation function, we were able to Fourier transform the expression and compare it with the previously calculated Fourier-space result.  We then developed a numerical simulation of the Zel'dovich approximation in Python whose results agreed with the analytical expression. We developed the numerical code in order to test future results, where analytical comparisons are not available to us, such as the redshift-space result presented in this paper. Here, we rely solely on these numerical simulations of the Zel'dovich approximation to verify the result.

\begin{figure}
\begin{center}
\includegraphics[width=0.5\textwidth]{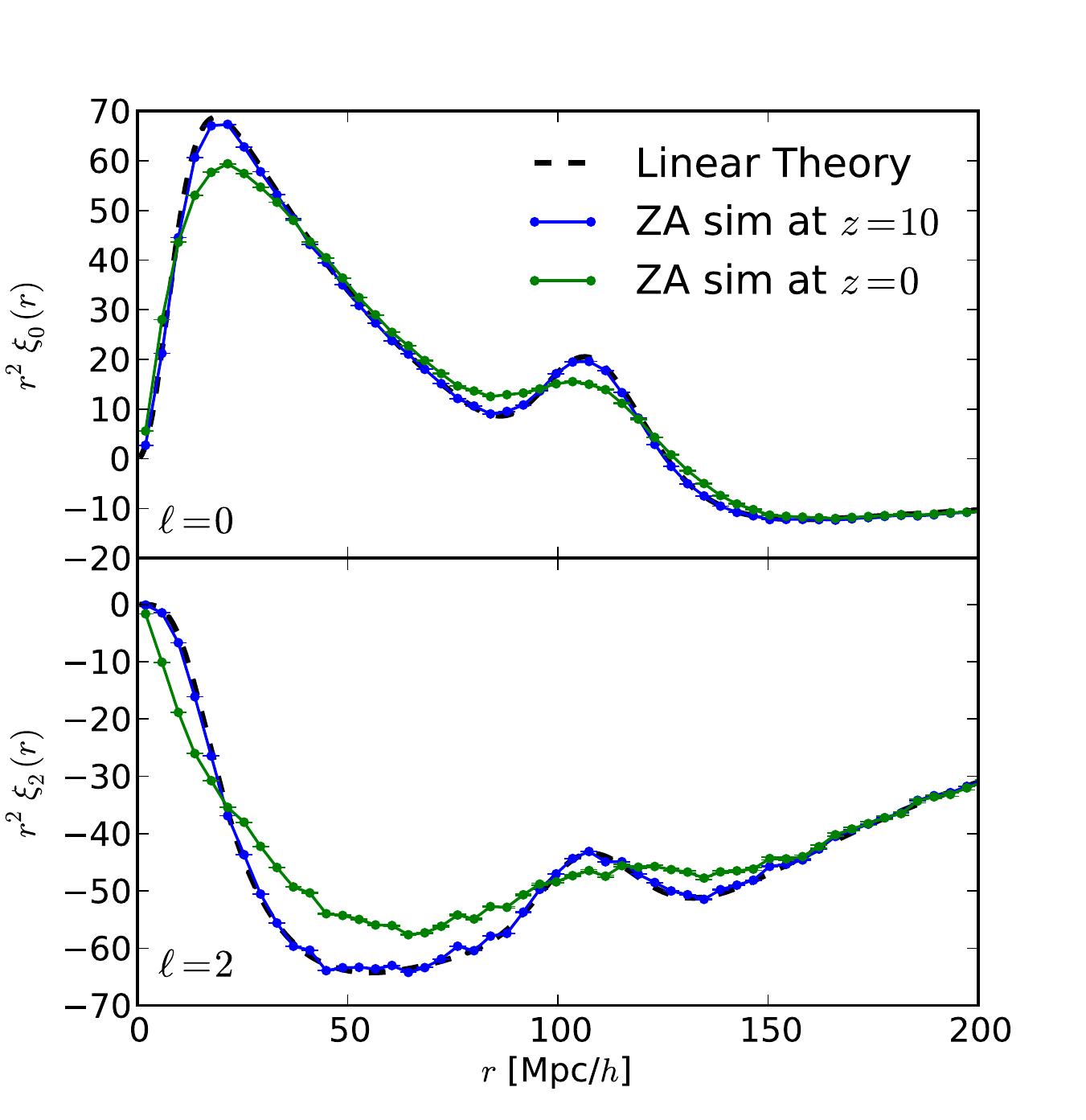}
\caption{Top: Monopole of ZA sims at $z=10$ and $z=0$ vs linear theory Bottom: quadrupole. The standard error bars from the 600 realizations are too small to be shown.}
\label{fig:zafullpoles}
\end{center}
\end{figure}

\begin{figure*}
\begin{center}
\begin{subfloat}[$z=10$]{
\includegraphics[scale=0.43]{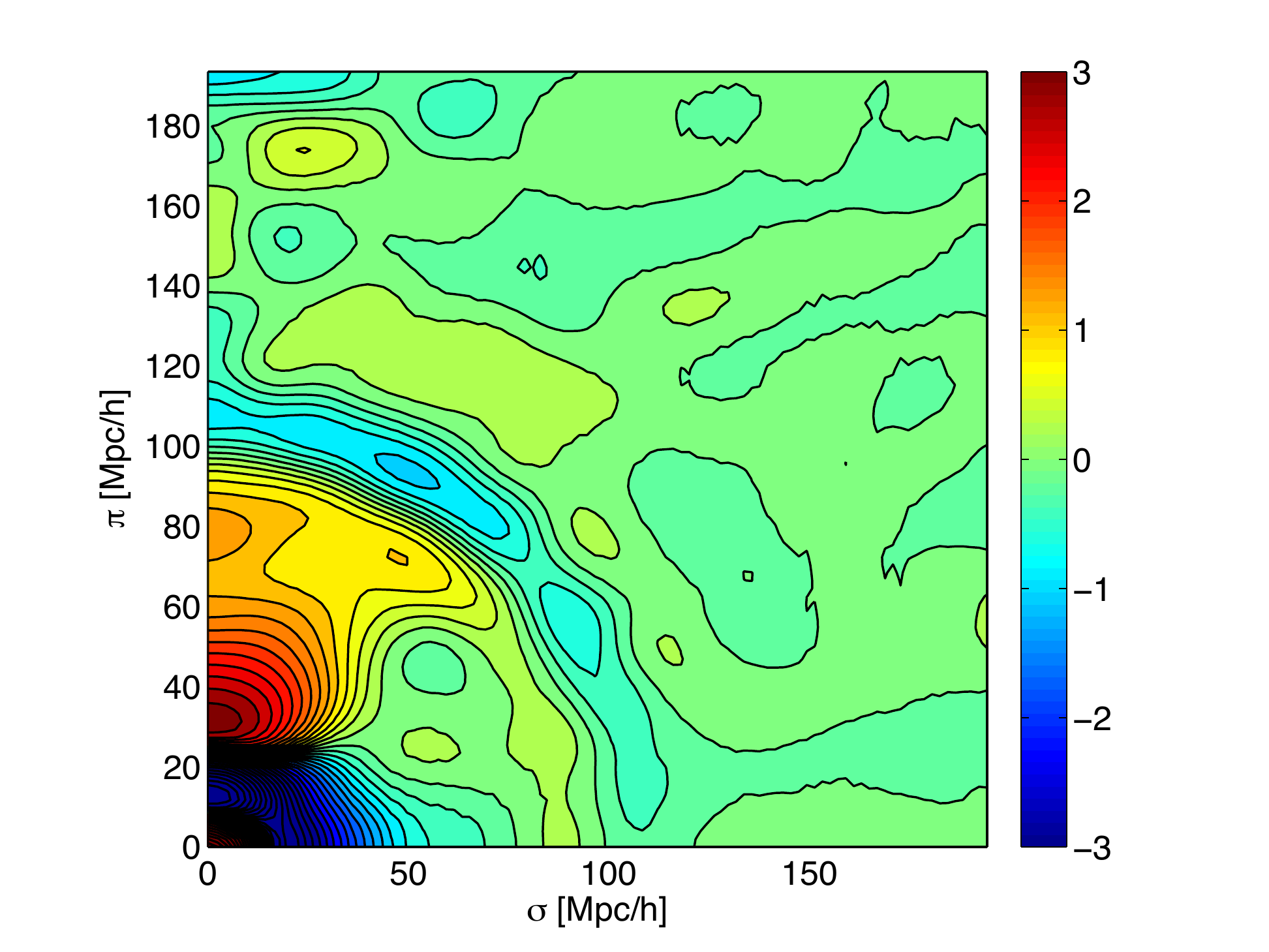}}
\end{subfloat}
\begin{subfloat}[$z=5$]{
\includegraphics[scale=0.43]{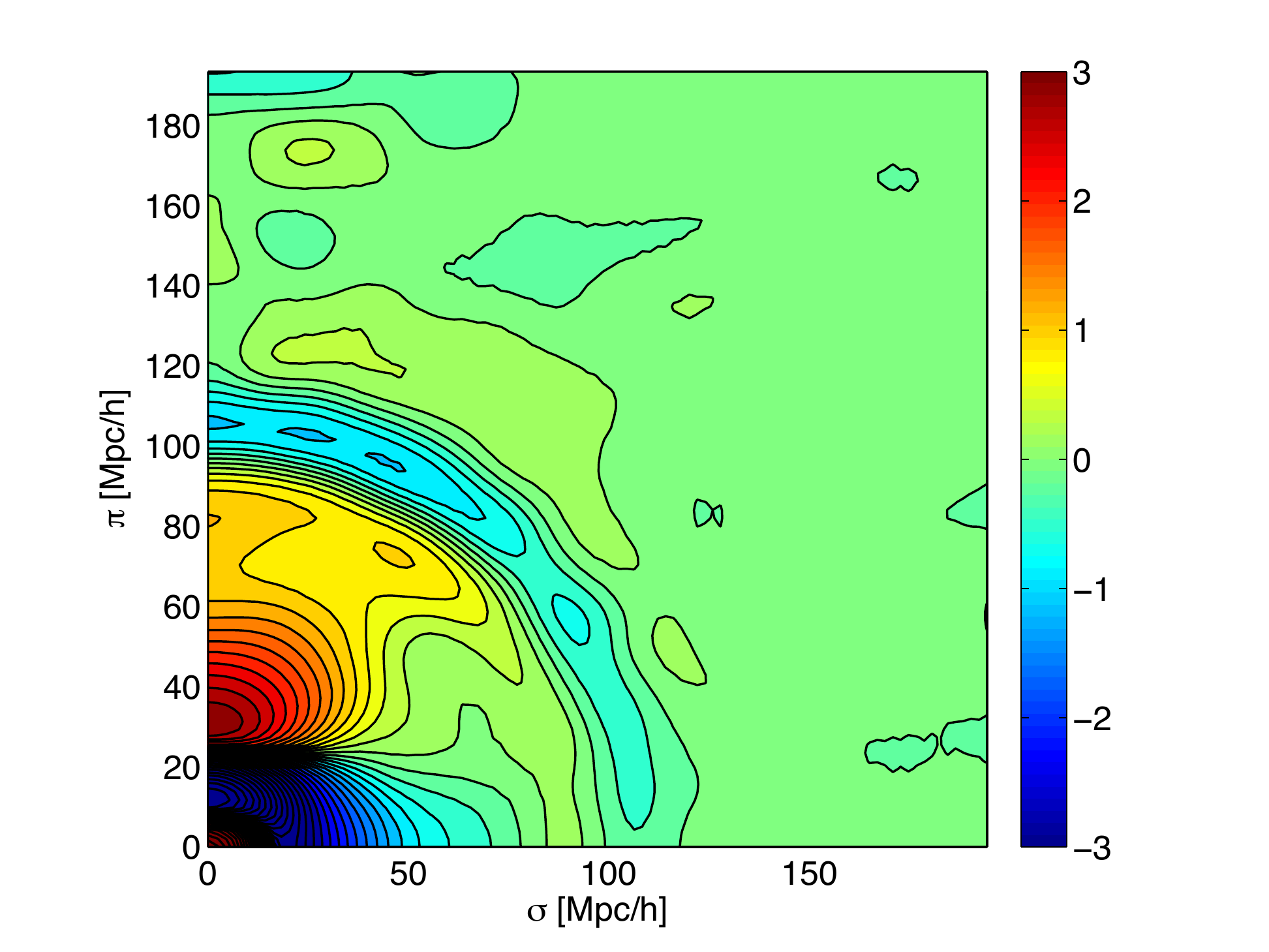}}
\end{subfloat}
\\
\begin{subfloat}[$z=1$]{
\includegraphics[scale=0.43]{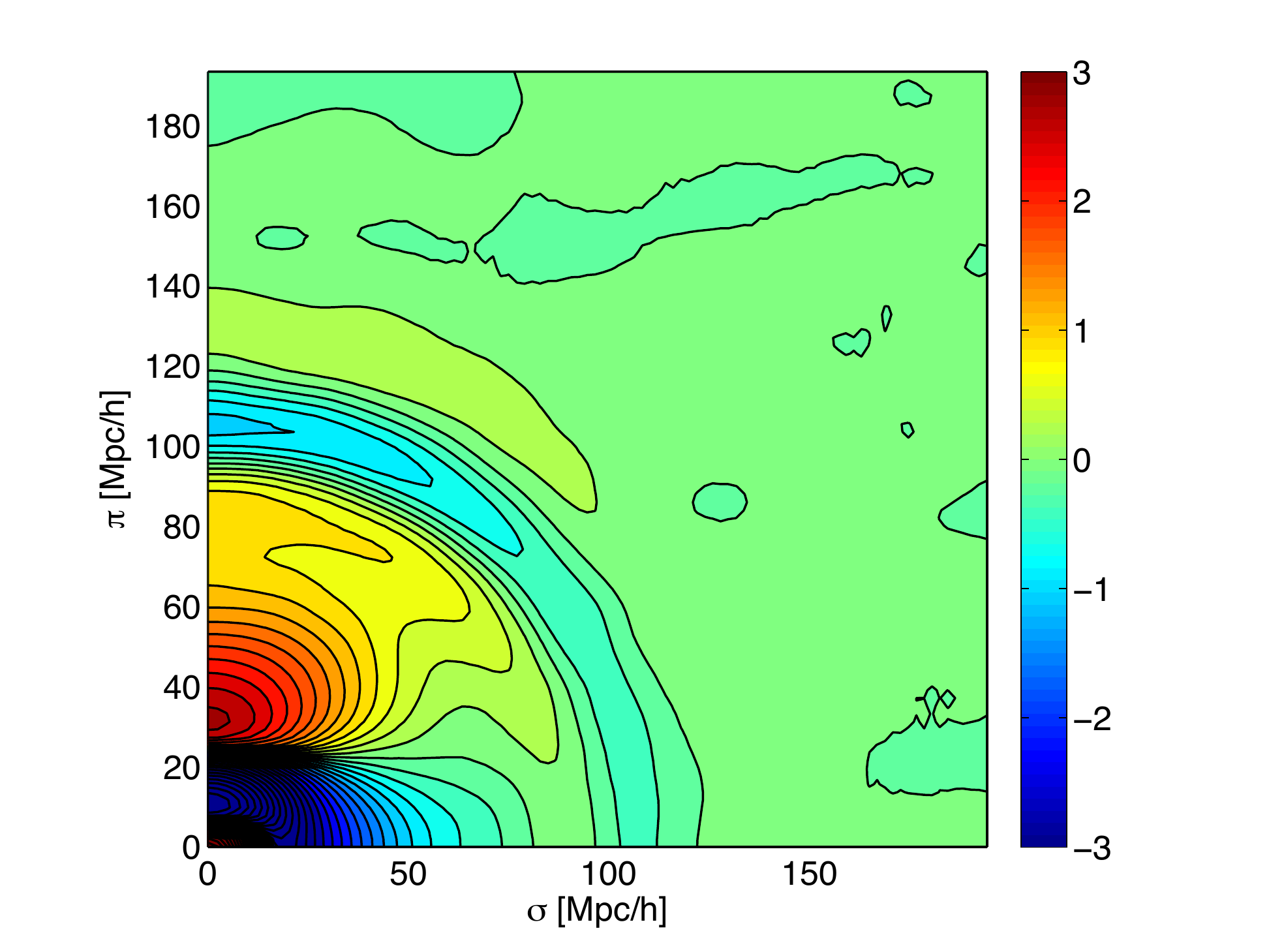}}
\end{subfloat}
\begin{subfloat}[$z=0$]{
\includegraphics[scale=0.43]{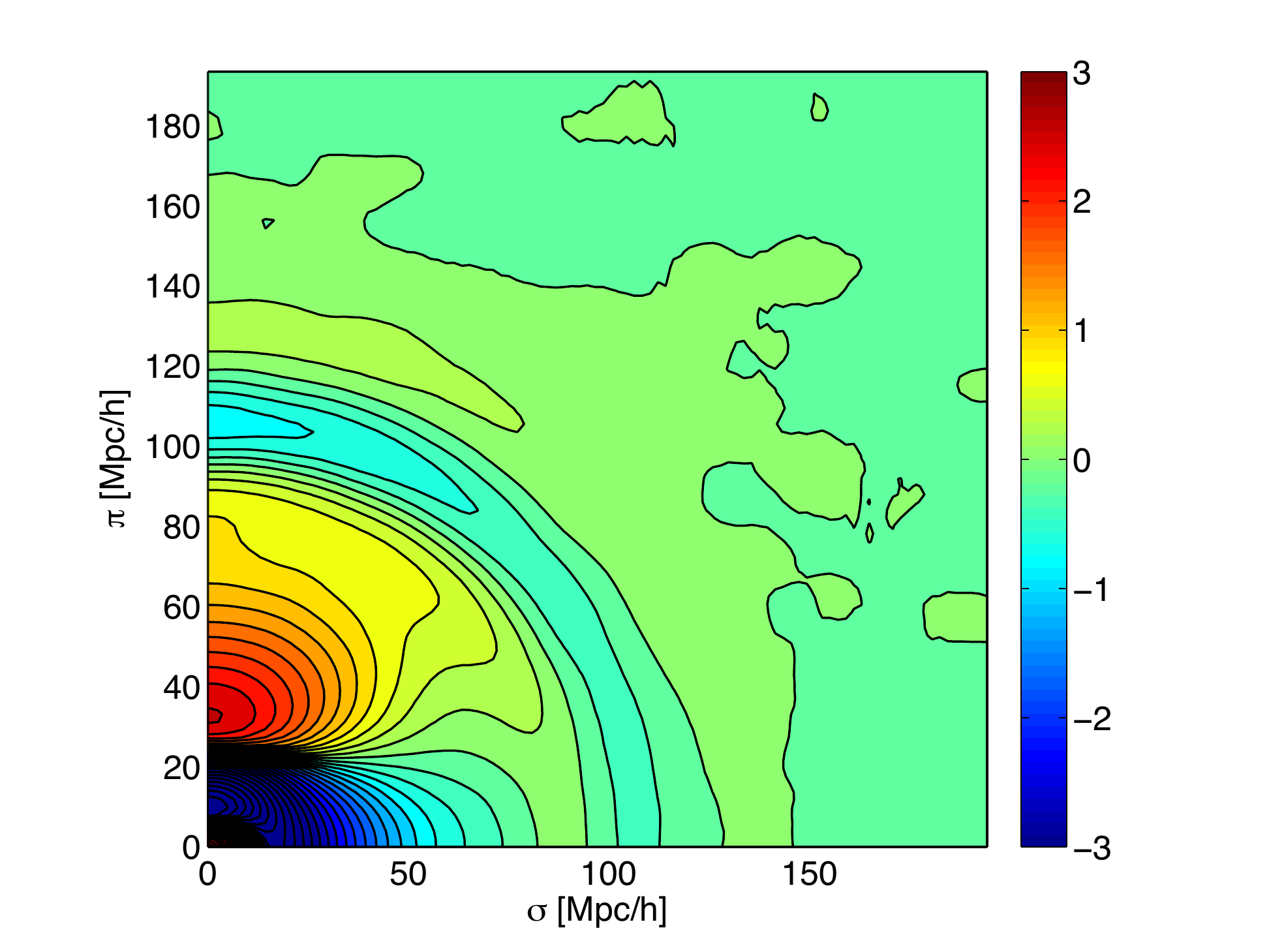}}
\end{subfloat}
\caption{Nonlinear correlation functions from 1000 Zel'dovich simulations, at $z=10$ (top left), $z=5$ (top right), $z=1$ (bottom left), and $z=0$ (bottom right).}
\label{fig:cfnonlinsims}
\end{center}
\end{figure*}

We use {\scshape camb} \citep{camb} to generate the initial power spectrum, with the cosmological parameters $\Omega_{\Lambda}=0.71$, $\Omega_m=0.29$, $\Omega_b= 0.045$, $h=0.7$, and $\sigma_8(z=0)=0.89$. Each realization is a $1$ Gpc/h box with $1024^3$ particles. The particles are displaced from the initial grid to their redshift-space positions at $z=10, 5, 1, 0$.  We calculate the final redshift-space density using a cloud-in-cell scheme on a $512^3$ grid. 

To compute the 2-dimensional redshift-space correlation function, we first FFT the density grid, square it to get the 3-dimensional power spectrum, and then inverse FFT the result. We compute averages along transverse slices of the 3-dimensional correlation function in order to find the 2-dimensional $\xi(\sigma, \pi)$ in a given realization. The average of the 2-dimensional correlation functions over $600$ realizations at each redshift allows us to study the behavior of the correlation function over time.

First, we consider the analytic prediction for the correlation function. Figure \ref{fig:cflin} shows the redshift-space correlation function expected from linear theory, Equation \ref{eq:cflin}. The $x$-axis is the transverse direction ($\sigma$), and the $y$-axis is the line-of-sight direction ($\pi$). The BAO feature shows up as an anisotropic arc in the 2-dimensional redshift-space correlation function. 

Figure \ref{fig:cfnonlin} shows the prediction for the nonlinear correction term from the Zel'dovich approximation, Equation \ref{eq:xi2rs}. The negative value along the BAO arc indicates that this term dampens the BAO feature. Along the line of sight, this term also tends to broaden the arc, due to the positive contribution on either side of the peak. 

Next, we look at the results from the Zel'dovich simulations. A useful way to study the the redshift-space correlation function is through its multipole moments. The linear term, Equation \ref{eq:cflin}, contains multipoles $\ell=0$, $\ell=2$, and $\ell=4$. The nonlinear term contains even poles up to $\ell=8$. Figure \ref{fig:zafullpoles} shows the monopole and quadrupole of the measured redshift-space correlation functions at $z=10$ and $z=0$ compared to the prediction from linear theory. The correlation functions have been normalized by $D(z)^2$ so they lie on top of each other. As expected, the correlation function at $z=10$ agrees very well with linear theory. This agreement breaks down at $z=0$, where nonlinear evolution has damped the BAO feature significantly compared to the linear theory prediction. From this figure we can see that the amplitude of the nonlinear correction at $z=0$ is about 10\%-20\% at the BAO peak in the monopole and quadrupole terms.

We would like to test the validity of our analytic expression for the nonlinear correction, Equation \ref{eq:xi2rs}. To directly compare the simulations to our expression for the nonlinear term, we subtract off the linear term at each redshift and only look at the difference. Figure \ref{fig:cfnonlinsims} shows the average nonlinear contribution to the correlation functions at redshifts $z=10, 5, 1, 0$ from the Zel'dovich simulations. We have divided each by $D(z)^4$ so they are all on the same scale. Note that at high redshift, what is shown is the difference between two already small values (the measured and linear correlation functions), divided by $D(z)^4$, which is also very small. Thus, the signal at high redshifts is noisier, even when averaged over 600 simulations. Window function and discreteness effects are also amplified. At $z=0$, where the nonlinear signal is larger, higher-order contributions become non-negligible, thus agreement with our theoretical expression is not guaranteed. At $z=1$ we see the best agreement with our theoretical prediction, because the nonlinear term is larger, and higher-order terms are still small. We compare the results from the simulations at various redshifts in Figure \ref{fig:cfnonlinsims} to the analytical expression in Figure \ref{fig:cfnonlin}. The shape of the nonlinear term matches well to the analytical result at all redshifts. The damping of the BAO feature and broadening along the line of sight are clear.

\begin{figure}
\begin{center}
\includegraphics[width=0.5\textwidth]{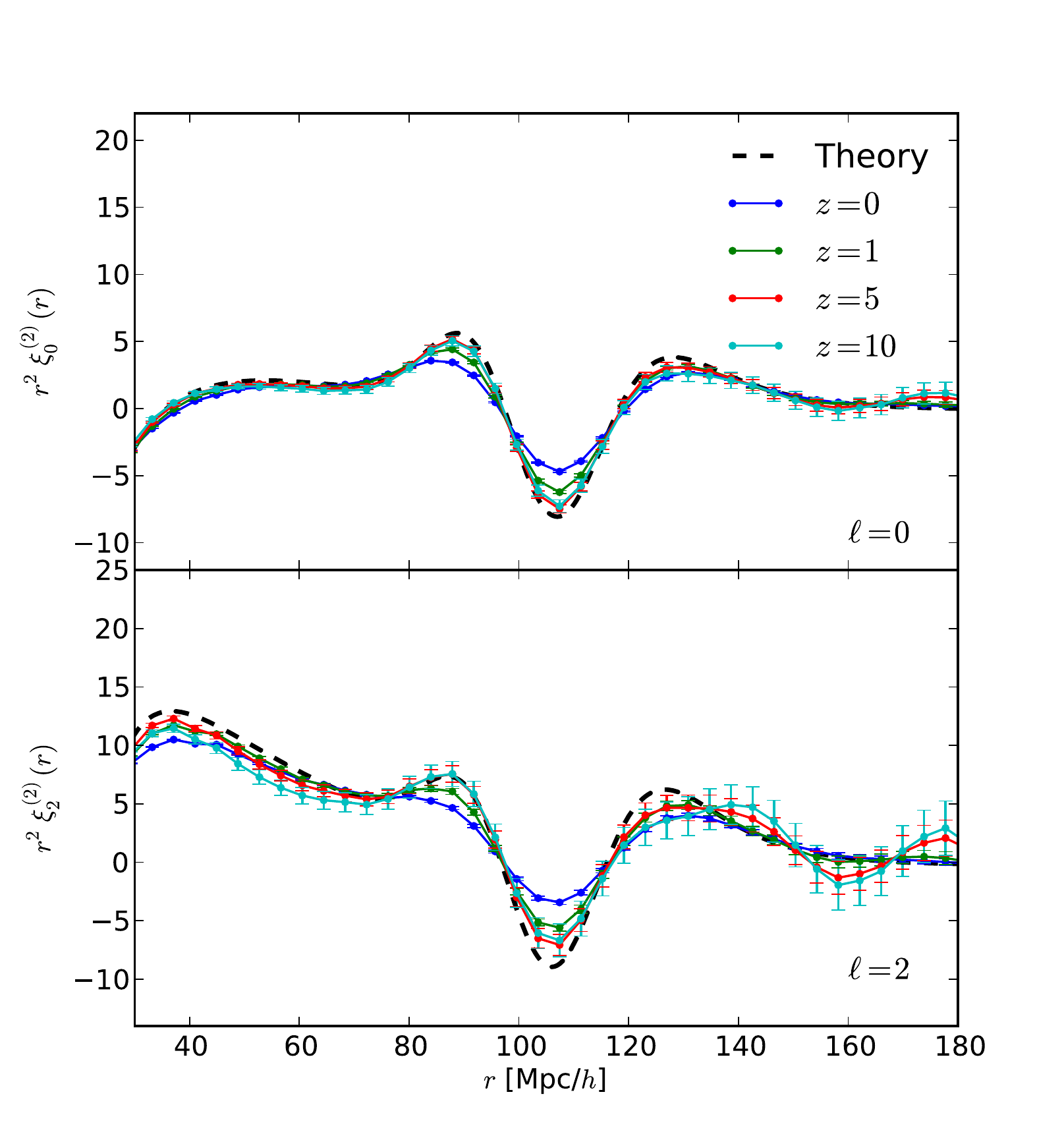}
\caption{Top: the monopole of nonlinear correction term measured from Zel'dovich realizations (points) versus the analytic prediction (black dashed line). Bottom: the quadrupole of nonlinear correction term. Error bars show the standard error from 600 realizations. The larger error bars at high redshift is due to the normalization by $D(z)^4$.}
\label{fig:simpoles}
\end{center}
\end{figure}

To more easily compare the numerical results to the analytical expression, we look at the monopole and quadrupole moments of the nonlinear correction term. Figure \ref{fig:simpoles} shows the monopole and quadrupole moments of the analytic expression, Equation \ref{eq:xi2rs}, compared to those measured from the Zel'dovich realizations. The multipoles are estimated numerically by performing azimuthal averages over the 2-dimensional correlation function, weighted by the corresponding Legendre polynomial. Measuring poles higher than $\ell=2$ in this way is limited by the resolution of the grid, since the poles become increasingly oscillatory. However, we can see that the first two poles of the nonlinear term agree well with the analytic prediction, especially at high redshift. At low redshift, the contribution from higher-order terms is not negligible, and thus the agreement is not perfect. The overall agreement between the numerical and analytical results allows us to conclude that our expression for the nonlinear term in redshift space is correct.

Other recent work using the Zel'dovich approximation to model the nonlinear evolution of matter, such as \citet{white2014, carlson2013, tassev2014a, tassev2014b}, show similar results. The Zel'dovich expression computed in \citet{carlson2013} (Equation 34) and used in \citet{white2014} gives the full Zel'dovich correlation function in closed form, involving a numerical integral of a Gaussian function. This result would presumably agree exactly with our numerical Zel'dovich realizations at all redshifts. 

In contrast, the result presented here is a perturbative expansion of the full Zel'dovich approximation, giving the equivalent of the Standard Perturbation Theory 1-loop correlation function for the Zel'dovich approximation. This result is computationally simpler than the full Zel'dovich expression, involving only products of linear correlation functions ($\xi_n^m(r)$), which are 1-dimensional integrals and can be tabulated easily for a given initial power spectrum. As with Standard Perturbation Theory, the 1-loop correlation function does not exactly predict the low-redshift behavior of the full Zel'dovich approximation due to higher-order terms becoming non-negligible, as seen in Figure \ref{fig:simpoles}.

We now compare our result to full $N$-body simulations. We use the Indra suite of simulations, which consists of 100 dark-matter only simulations run with the Gadget code \citep{indra}. Each realization is a $(1 \text{Gpc}/h)^3$ box with $1024^3$ particles and WMAP 9 parameters \citep{wmap9}. We compute the redshift-space correlation functions as described above using a $1024^3$ CIC scheme for the density. 

\begin{figure}
\begin{center}
\includegraphics[width=0.5\textwidth]{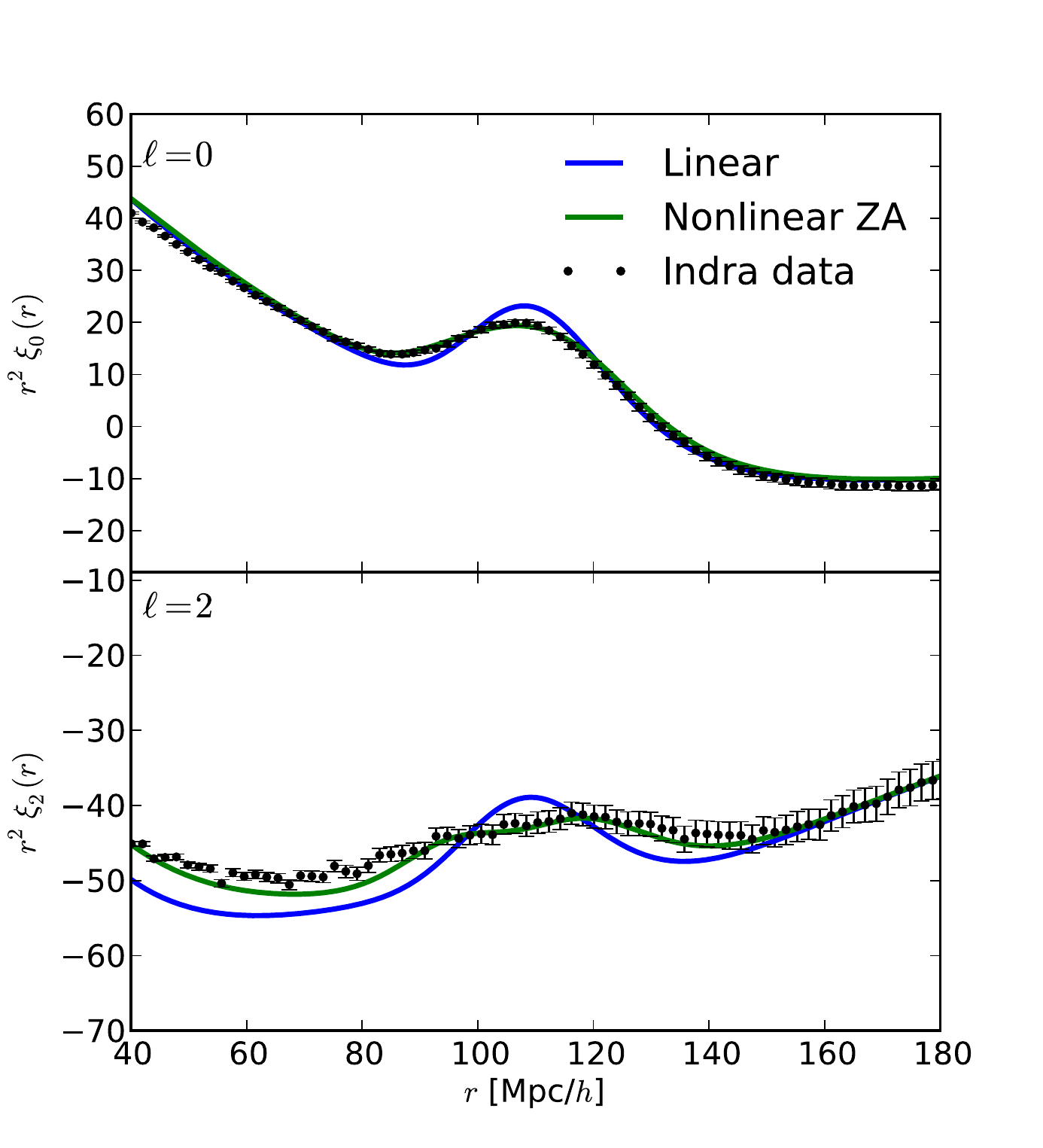}
\caption{Top: Monopole moment of Indra simulations at $z=0$ (black points) versus linear theory (blue line) and 1-loop Zel'dovich prediction (green line). Bottom: quadrupole moment. Error bars are standard deviation in the 100 Indra realizations.}
\label{fig:zavsindra}
\end{center}
\end{figure}

Figure \ref{fig:zavsindra} shows the monopole and quadrupole moments of the measured redshift-space correlation functions at $z=0$ compared to the predictions from linear theory and our perturbative expansion of the Zeldovich approximation. The error bars show the standard error of the multipoles measured in the 100 realizations. The results from these $N$-body simulations agree quite well with our analytic expression on BAO scales at low redshift. This indicates that the 1-loop Zel'dovich correlation function may be sufficient for modeling the BAO peak at low redshift. One application of our result is that it could be used as a template for determining the BAO peak location.

\section{Conclusion}
\label{sec:conclusion}

We have shown that our configuration-space approach to perturbation theory can be extended to redshift space to give the nonlinear redshift-space correlation function to 2nd order in the Zel'dovich approximation. We validate our result through comparison with numerical simulations of the full Zel'dovich approximation. While the expression for the nonlinear contribution is made up of many terms, it is computationally simple because it is a closed-form expression made up of products of linear correlation functions, $\xi_n^m(s)$. 

We then compared our analytical model to results from full $N$-body simulations. We showed that it predicts the first two multipole moments of the redshift-space correlation function quite well on BAO scales at $z=0$. Due to the numerical simplicity of the result, it may be useful as a template for constraining cosmological parameters from BAO measurements. 

In principle, this method can be extended both to higher orders in the Zel'dovich approximation, beyond 2nd order, and to higher orders in Lagrangian Perturbation Theory, beyond the Zel'dovich approximation. Beyond 2nd order, the terms will be made up of more products of linear correlation functions, and should be straightforward to compute. Extending this method to higher orders in Lagrangian Perturbation Theory would result in more accurate results, comparable to standard perturbation theory. We leave these extensions to future work.

\acknowledgements
We thank Mark Neyrinck and Xin Wang for many useful discussions. This research has been supported by the Gordon and Betty Moore Foundation and NSF OIA grant CDI-1124403.

\appendix
The non-vanishing coefficients $B_{\ell}(f)$ and $A_{n_1, n_2}^{\ell, m}(f)$ are defined below, grouped by multipole $\ell$:
\raggedright
\resizebox{7in}{!}{
\begin{tabular}{m{3.6in}  m{3.6in}}
$B_0=-\frac{1}{105}\left(105+140 f+140 f^2+72 f^3+15 f^4\right)$ & $A_{0,0}^{0,0}=\frac{1}{225} \left(285+380 f+320 f^2+144 f^3+27 f^4\right)$ \\[7pt]
$A_{2,2}^{0, 0}=\frac{2}{2205}\left(1785+2380 f+2233 f^2+1098 f^3+216 f^4\right)$ & $A_{4,4}^{0,0}=\frac{4 }{3675}\left(105+140 f+140 f^2+72 f^3+16 f^4\right)$ \\[7pt]
$A_{1,1}^{0, 1}=-\frac{4}{525} \left(420+560 f+497 f^2+234 f^3+45 f^4\right)$ & $A_{3, 3}^{0, 1}=-\frac{4 }{1575}\left(315+420 f+399 f^2+198 f^3+40 f^4\right)$ \\[7pt]
$A_{0,0}^{0, 2}=\frac{1}{315}\left(105+140 f+140 f^2+72 f^3+15 f^4\right)$ & $A_{2,2}^{0, 2}=\frac{2}{315} \left(105+140 f+119 f^2+54 f^3+10 f^4\right)$ \\[7pt] \cline{1-2}\\
$B_2=\frac{2}{735}  \left(980+1295 f+738 f^2+159 f^3\right)$ & $A_{0, 2}^{2, 0}=-\frac{4}{315}  \left(385+484 f+261 f^2+54 f^3\right)$\\[7pt]
$A_{2,2}^{2,0}=\frac{4 }{3087} \left(1309+1786 f+1017 f^2+216 f^3\right)$ & $A_{2,4}^{2, 0}=-\frac{16 }{1715} \left(140+179 f+106 f^2+24 f^3\right)$\\[7pt]
 $A_{4,4}^{2, 0}=\frac{16}{11319}  \left(77+110 f+66 f^2+16 f^3\right)$ & $A_{1,1}^{2, 1}=-\frac{8}{525}  \left(308+395 f+216 f^2+45 f^3\right)$ \\[7pt]
$A_{1,3}^{2, 1}=\frac{4 }{1225}\left(581+740 f+417 f^2+90 f^3\right)$ & $A_{3,1}^{2, 1}=\frac{4}{525} \left(399+510 f+283 f^2+60 f^3\right)$ \\[7pt]
$A_{3,3}^{2, 1}=-\frac{16 }{4725}\left(231+315 f+182 f^2+40 f^3\right)$ & $A_{5,3}^{2, 1}=\frac{16 }{14553} \left(231+297 f+176 f^2+40 f^3\right)$ \\[7pt]
$A_{0,2}^{2, 2}=-\frac{2}{315}  \left(98+131 f+72 f^2+15 f^3\right)$ & $A_{2,0}^{2,2}=-\frac{2}{63} \left(28+37 f+22 f^2+5 f^3\right)$ \\[7pt]
$A_{2,2}^{2,2}=\frac{4}{441} \left(77+92 f+49 f^2+10 f^3\right)$ & $A_{4,2}^{2,2}=-\frac{16 }{8085} \left(231+297 f+154 f^2+30 f^3\right)$ \\[7pt]\cline{1-2}\\
$B_4=-\frac{8}{245}  \left(21+19 f+6 f^2\right)$ & $A_{0,2}^{4, 0}=\frac{8}{175} f^2 \left(25+22 f+6 f^2\right)$\\[7pt]
$A_{2,2}^{4,0}=\frac{72 }{1715}\left(51+46 f+12 f^2\right)$ & $A_{2,4}^{4, 0}=-\frac{32 }{3773} \left(88+85 f+24 f^2\right)$ \\[7pt]
$A_{4,4}^{4, 0}=\frac{144}{1226225}  \left(715+702 f+216 f^2\right)$ &  $A_{1,3}^{4, 1}=\frac{48 }{1225} \left(41+37 f+10 f^2\right)$\\[7pt]
$A_{3,1}^{4, 1}=\frac{16 }{1575} \left(251+227 f+60 f^2\right)$ &$A_{3,3}^{4, 1}=-\frac{16 }{5775} \left(231+217 f+60 f^2\right)$\\[7pt]
$A_{5,1}^{4, 1}=-\frac{16}{3465}\left(121+109 f+30 f^2\right)$ &$A_{5,3}^{4, 1}=\frac{32 }{105105}\left(429+416 f+120 f^2\right)$ \\[7pt]
$A_{2,2}^{4, 2}=\frac{8}{735} \left(61+56 f+15 f^2\right)$ &$A_{4,0}^{4, 2}=\frac{8 }{1155}\left(55+52 f+15 f^2\right)$\\[7pt]
$A_{4,2}^{4, 2}=-\frac{32 }{17787}\left(143+119 f+30 f^2\right)$&$A_{6,2}^{4, 2}=\frac{16 }{33033}  \left(143+130 f+30 f^2\right)$ \\[7pt]\cline{1-2}\\
$A_{2,4}^{6, 0}=-\frac{32}{539}  (13+6 f)$ &$A_{4,4}^{6, 0}=\frac{32 }{8085} (15+8 f)$\\[7pt]
$A_{3,3}^{6, 1}=-\frac{32 }{2079}(43+20 f)$&$A_{5,1}^{6, 1}=-\frac{32 }{1155} (13+6 f)$\\[7pt]
$A_{5,3}^{6, 1}=\frac{64 }{1485} (2+f)$&$A_{4,2}^{6, 2}=-\frac{16}{2541} (32+15 f)$\\[7pt]
$A_{6,0}^{6, 2}=-\frac{16}{693}  (2+f)$ & $A_{6,2}^{6, 2}=\frac{32 }{5445}(5+2 f)$ \\[7pt]\cline{1-2}\\
$A_{4,4}^{8, 0}=\frac{128}{2145}$ & $A_{5,3}^{8, 1}=\frac{512 }{6435}$ \\[7pt]
$A_{6,2}^{8, 2}=\frac{128 }{6435}$&\\[7pt]\cline{1-2}\\
\end{tabular}
}

The nonlinear contribution to the monopole ($\ell=0$) term is:
\begin{align}
\bar{\xi}^{(2)}_0(s)&=\left(\frac{19}{15}+\frac{76 f}{45}+\frac{64 f^2}{45}+\frac{16 f^3}{25}+\frac{3 f^4}{25}\right) \xi_0^0(s)^2
+\left(\frac{1}{3}+\frac{4 f}{9}+\frac{4 f^2}{9}+\frac{8 f^3}{35}+\frac{f^4}{21}\right) \xi_0^{2}(s) \left(\xi_0^{-2}(s)-3\sigma_v^2\right)\notag\\
&-\left(\frac{16}{5}+\frac{64 f}{15}+\frac{284 f^2}{75}+\frac{312 f^3}{175}+\frac{12 f^4}{35}\right) \xi_1^{-1}(s) \xi_1^1(s)+\left(\frac{34}{21}+\frac{136 f}{63}+\frac{638 f^2}{315}+\frac{244 f^3}{245}+\frac{48 f^4}{245}\right) \xi_2^0(s)^2\notag\\
&+\left(\frac{2}{3}+\frac{8 f}{9}+\frac{34 f^2}{45}+\frac{12 f^3}{35}+\frac{4 f^4}{63}\right) \xi_2^{-2}(s) \xi_2^2(s)-\left(\frac{4}{5}+\frac{16 f}{15}+\frac{76 f^2}{75}+\frac{88 f^3}{175}+\frac{32 f^4}{315}\right) \xi_3^{-1}(s) \xi_3^1(s)\notag\\
&+\left(\frac{4}{35}+\frac{16 f}{105}+\frac{16 f^2}{105}+\frac{96 f^3}{1225}+\frac{64 f^4}{3675}\right) \xi_4^0(s)^2\notag
\end{align}

The nonlinear contribution to the quadrupole ($\ell=2$) term is:
\begin{align}
\bar{\xi}^{(2)}_2(s)&=-\left(\frac{28 f}{225}+\frac{262 f^2}{1575}+\frac{16 f^3}{175}+\frac{2 f^4}{105}\right) \xi_0^2(s) \xi_2^{-2}(s)-\left(\frac{44 f}{45}+\frac{1936 f^2}{1575}+\frac{116 f^3}{175}+\frac{24 f^4}{175}\right) \xi_0^0(s) \xi_2^0(s)\notag\\
&+\left(\frac{748 f}{2205}+\frac{7144 f^2}{15435}+\frac{452 f^3}{1715}+\frac{96 f^4}{1715}\right) \xi_2^0(s)^2+\left(\frac{8 f}{45}+\frac{74 f^2}{315}+\frac{164 f^3}{1225}+\frac{106 f^4}{3675}\right)\xi_2^2(s)\left(3\sigma_v^2-\xi_2^{-2}(s)\right)\notag\\
&+\xi_1^{-1}(s) \left(\left(\frac{76 f}{125}+\frac{136 f^2}{175}+\frac{1132 f^3}{2625}+\frac{16 f^4}{175}\right) \xi_3^1(s) -\left(\frac{352 f}{375}+\frac{632 f^2}{525}+\frac{576 f^3}{875}+\frac{24 f^4}{175}\right) \xi_1^1(s)\right)\notag\\
&+\left(\frac{332 f}{875}+\frac{592 f^2}{1225}+\frac{1668 f^3}{6125}+\frac{72 f^4}{1225}\right) \xi_1^1(s) \xi_3^{-1}(s)-\left(\frac{64 f}{245}+\frac{2864 f^2}{8575}+\frac{1696 f^3}{8575}+\frac{384 f^4}{8575}\right) \xi_2^0(s) \xi_4^0(s)\notag\\
&+\xi_2^{-2}(s) \left(\left(\frac{44 f}{315}+\frac{368 f^2}{2205}+\frac{4 f^3}{45}+\frac{8 f^4}{441}\right) \xi_2^2(s)-\left(\frac{16 f}{175}+\frac{144 f^2}{1225}+\frac{32 f^3}{525}+\frac{32 f^4}{2695}\right) \xi_4^2(s)\right)\notag\\
&+\xi_3^{-1}(s) \left(\left(\frac{16 f}{315}+\frac{16 f^2}{245}+\frac{256 f^3}{6615}+\frac{128 f^4}{14553}\right) \xi_5^1(s)-\left(\frac{176 f}{1125}+\frac{16 f^2}{75}+\frac{416 f^3}{3375}+\frac{128 f^4}{4725}\right) \xi_3^1(s)\right)\notag\\
&+\left(\frac{16 f}{735}+\frac{32 f^2}{1029}+\frac{32 f^3}{1715}+\frac{256 f^4}{56595}\right) \xi_4^0(s)^2\notag
\end{align}

\bibliographystyle{apj}
\bibliography{zeldptbib}

\end{document}